\documentclass[letterpaper,twocolumn,conference]{IEEEtran}
\usepackage[T1]{fontenc}
\usepackage{color}
\usepackage{array}
\usepackage{multirow}
\usepackage{amsmath}
\usepackage{amssymb}
\usepackage{graphicx}
\usepackage{algpseudocode}
\usepackage{algorithm}
\usepackage[unicode=true,
 bookmarks=true,bookmarksnumbered=true,bookmarksopen=true,bookmarksopenlevel=1,
 breaklinks=false,pdfborder={0 0 0},backref=false,colorlinks=false]
 {hyperref}
\hypersetup{pdftitle={Your Title},
 pdfauthor={Your Name},
 pdfpagelayout=OneColumn, pdfnewwindow=true, pdfstartview=XYZ, plainpages=false}

\makeatletter

\pdfpageheight\paperheight
\pdfpagewidth\paperwidth

\providecommand{\tabularnewline}{\\}

\usepackage[caption=false,font=footnotesize]{subfig}
\usepackage[square, comma, sort&compress, numbers]{natbib}

\makeatother

\begin{document}

\title{STARIMA-based Traffic Prediction with Time-varying Lags}

\author{\IEEEauthorblockN{Peibo Duan\IEEEauthorrefmark{1}, Guoqiang Mao\IEEEauthorrefmark{1}\IEEEauthorrefmark{2}\IEEEauthorrefmark{3}\IEEEauthorrefmark{4},
Shangbo Wang\IEEEauthorrefmark{1}, Changsheng Zhang\IEEEauthorrefmark{5}
and Bin Zhang\IEEEauthorrefmark{5}}\IEEEauthorblockA{\IEEEauthorrefmark{1}School of Computing and Communications\\
University of Technology, Sydney, Australia}\IEEEauthorblockA{\IEEEauthorrefmark{2}Data61 Australia}\IEEEauthorblockA{\IEEEauthorrefmark{4}Beijing Unriversity of Posts and Telecommunications,
Beijing, China}\IEEEauthorblockA{\IEEEauthorrefmark{4}School of Electronic Information Communications,
Huazhong Unriversity of Science Technology, Wuhan, China}\IEEEauthorblockA{\IEEEauthorrefmark{5}School of Computer Application\\
Northeastern University\\
Bin Zhang is the correspondence author }}
\maketitle
\begin{abstract}
Based on the observation that the correlation between observed traffic
at two measurement points or traffic stations may be time-varying,
attributable to the time-varying speed which subsequently causes variations
in the time required to travel between the two points, in this paper,
we develop a modified Space-Time Autoregressive Integrated Moving
Average (STARIMA) model with time-varying lags for short-term traffic
flow prediction. Particularly, the temporal lags in the modified STARIMA
change with the time-varying speed at different time of the day or
equivalently change with the (time-varying) time required to travel
between two measurement points. Firstly, a technique is developed
to evaluate the temporal lag in the STARIMA model, where the temporal
lag is formulated as a function of the spatial lag (spatial distance)
and the average speed. Secondly, an unsupervised classification algorithm
based on ISODATA algorithm is designed to classify different time
periods of the day according to the variation of the speed. The classification
helps to determine the appropriate time lag to use in the STARIMA
model. Finally, a STARIMA-based model with time-varying lags is developed
for short-term traffic prediction. Experimental results using real
traffic data show that the developed STARIMA-based model with time-varying
lags has superior accuracy compared with its counterpart developed
using the traditional cross-correlation function and without employing
time-varying lags. 
\end{abstract}

\section{Introduction\label{sec:introduction}}

Road traffic prediction plays an important role in intelligent transport
systems by providing the required real-time information for traffic
management and congestion control, as well as the long-term traffic
trend for transport infrastructure planning \cite{2015_kkn,2015_urban_ARIMA,2014_short_term,2013_online_arima}.
Road traffic predictions can be broadly classified into short-term
traffic predictions and long-term traffic forecasts\cite{2013_short_term,2014_short_term,2015_short_term}.
Short-term prediction is essential for the development of efficient
traffic management and control systems, while long-term prediction
is mainly useful for road design and transport infrastructure planning.

There are two major categories of techniques for road traffic prediction:
those based on non-parametric models and those based on parametric
models. Non-parametric model based techniques, such as k-nearest neighbors
(KNN) model \cite{2015_kkn} and Artificial Neural Networks (ANN)
\cite{2012_ANN}, are inherently robust and valid under very weak
assumptions \cite{2002_comparison,2015_energy}, while parametric
model based techniques, such as auto-regressive integrated moving
average (ARIMA) model \cite{2006_application_ARIMA,2013_online_arima,2015_urban_ARIMA}
and its variants \cite{2009_STARIMA}\cite{2011_real_starima}, allows
to integrate knowledge of the underlying traffic process in the form
of traffic models that can then be used for traffic prediction. Both
categories of techniques have been widely used and in this paper,
we consider parametric model based techniques, particularly STARIMA
(Space-Time Autoregressive Integrated Moving Average)-based techniques.

As for the estimation of parameters and coefficients in STARIMA model,
overfitting easily occurs which makes the predictive performance poor
as it overreacts to minor fluctuations in the training data \cite{2002_book_statistic}.
Furthermore, the same model and hence the same correlation structure
is used for traffic prediction at different time of the day, which
is counter-intuitive and may not be accurate. To elaborate, consider
an artificial example of two traffic stations $A$ and $B$ on a highway,
where traffic station $B$ is at the down stream direction of $A$.
Intuitively, the correlation between the traffic observed at $A$
and the traffic observed at $B$ will peak at a time lag corresponding
to the time required to travel from $A$ to $B$ because at that time
lag, the (approximately) same set of vehicles that have passed $A$
now have reached $B$. Obviously, the time required to travel from
$A$ to $B$ depends on the traffic speed, which varies with the time
of the day, e.g. peak hours and off-peak hours. Accordingly, the time
lag corresponding to the peak correlation between the traffic at $A$
and the traffic at $B$ should also vary with time of the day and,
to be more specific, should approximately equal to the distance between
$A$ and $B$ divided by the mean speed of vehicles between $A$ and
$B$. Therefore, in designing the STARIMA model for traffic prediction,
the aforementioned time-varying lags should be taken into account
for accurate traffic prediction.

\begin{figure}[tbh]
\begin{centering}
\includegraphics[scale=0.34]{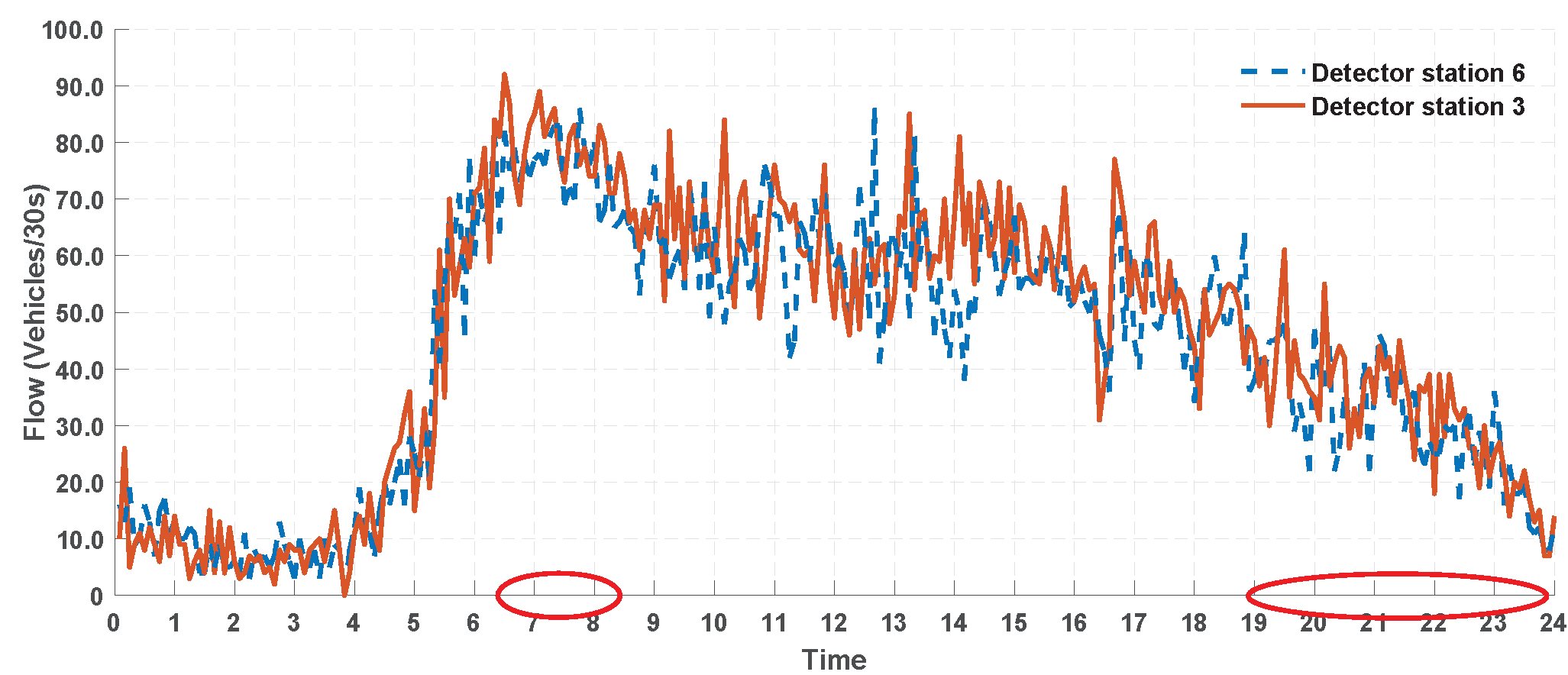}
\par\end{centering}

\caption{Traffic flow of station 6 in one day\label{fig:flow_3_6}}
\end{figure}

\begin{figure}[tbh]
\centering{}
\includegraphics[scale=0.42]{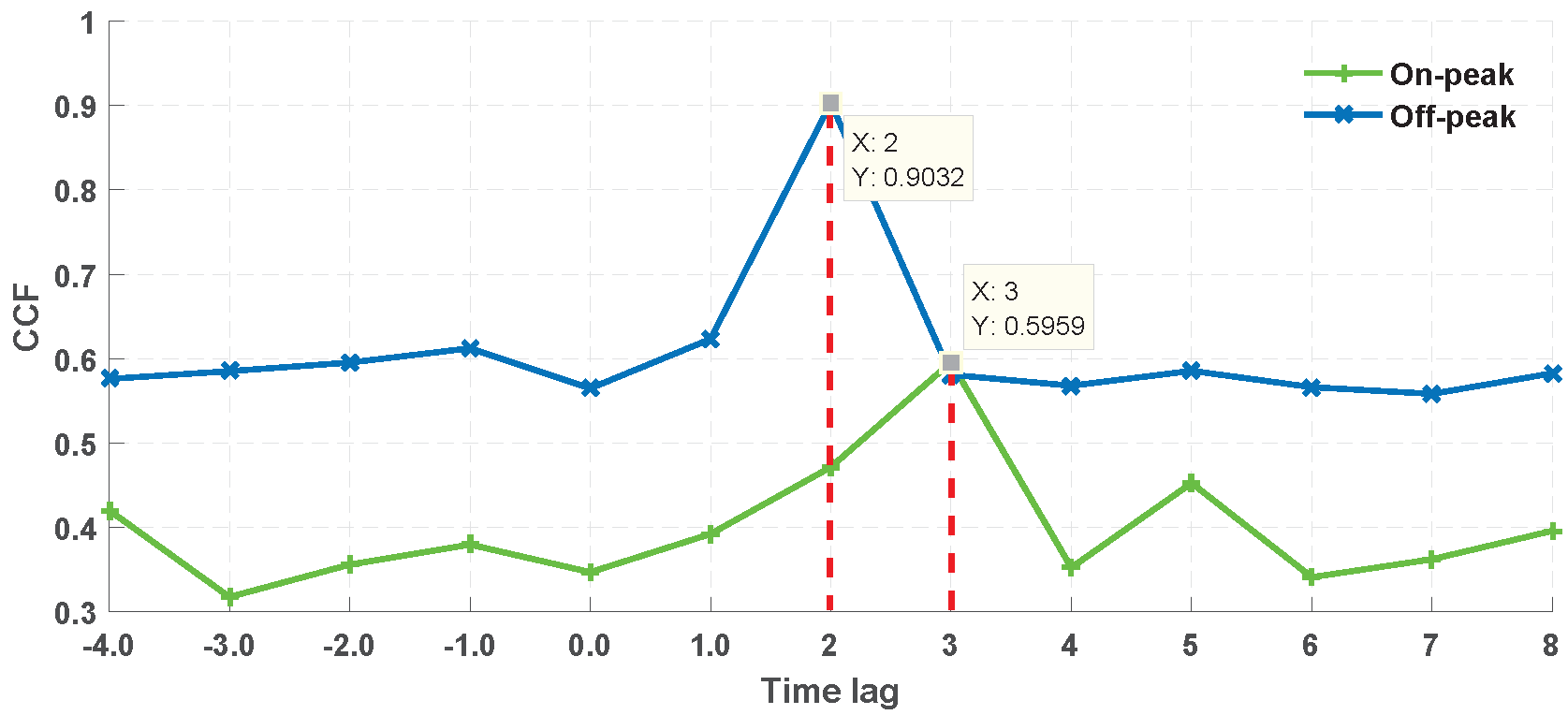}           
\caption{The CCF between traffic stations 6 and 3 in two different time periods\label{fig:CCF_3_6_different_time_period}}
\end{figure}

To validate the aforementioned intuition, we analyze the cross-correlation
function (CCF) of traffic flow data at two traffic stations (stations
6 and 3), denoted as $Corr_{63}$, from I-80 highway (more details
of data are discussed in Section \ref{sub:data_collection}) with
the formulation \eqref{eq:corr_u_y}:

\begin{equation}
Corr_{63}=\frac{E\left[(u_{t}-\bar{u})(y_{t+k}-\bar{y})\right]}{\sigma_{uu}\sigma{}_{yy}}\label{eq:corr_u_y}
\end{equation}
where $u$ and $y$ are the traffic flow data collected in $N$ time
slots from the two traffic stations, $k$ is the temporal order in
the range of $[0,1,2,...,N]\subset\mathbf{N}$, $\sigma_{uu}$ and
$\sigma_{yy}$ are respectively the standard deviation of $u$ and
$y$. A higher value of CCF indicates a stronger correlation of the
traffic at both stations. As shown in Fig.\ref{fig:flow_3_6}, the
correlation between traffic at stations 6 and 3 peaks at different
time lags depending on the time of the day. During on-peak period
(approximately from 6:30am - 8:30am), the correlation peaks at a time
lag of $3$ (one time lag corresponds to $30s$) while during off-peak
period (approximately from 19pm - 24pm), the correlation peaks at
a time lag of $2$, where one time lag corresponds to a time of 30$s$.
We observe that at peak hours, the time lag corresponding to the maximum
correlation is larger than that for off-peak hours. In the latter
section, we will further show that this time lag approximately equals
to the distance between the two traffic stations divided by the average
speed. Therefore, our intuition explained in the previous paragraph
is valid. 

The above observation motivates us to design a STARIMA-based traffic
prediction with time-varying lags which better matches the time-varying
correlation structure between traffic of different stations and hence
can potentially deliver more accurate traffic prediction. More specifically,
the contributions of the paper are:
\begin{itemize}
\item We analyze the CCF between the speed and traffic flow data between
different detector stations and establish the relationship between
the changes in the temporal lag (corresponding to the aforementioned
maximum correlation) and the speed variations.
\item An unsupervised classification algorithm based on ISODATA algorithm
is designed to classify different time periods of the day according
to the variation of the speed. The classification helps to determine
the appropriate time lag to use in the STARIMA model. 
\item A STARIMA-based model with time-varying lags is developed for short-term
traffic prediction. Experimental results using real traffic data show
that the developed STARIMA-based model with time-varying lags has
superior accuracy compared with its counterpart developed using the
traditional cross-correlation function and without employing time-varying
lags. 
\end{itemize}
The the rest of the paper is organized as follows. In Section \ref{sec:related_work},
we briefly discuss related work. Section \ref{sec:data_method} introduces
the STARIMA model and the ISODATA algorithm In Section \ref{sec:starima_time_lag_variation},
we present the details the proposed algorithm. The experimental results
are presented in Section \ref{sec:results}. Finally, Section \ref{sec:conclusion}
concludes the paper.

\section{Related Work\label{sec:related_work}}

There is previous work, which predicts traffic flow using a modified
ARIMA models \cite{2001_multivariate,2013_online_arima,2003_multivariate_ARIMA,2009_STARIMA,2011_real_starima}.
In \cite{2001_multivariate} and \cite{2003_multivariate_ARIMA},
a multivariate ARIMA based model, ARIMAX, was applied for better traffic
flow prediction. The difference is that the former paper considered
the varibility of the speed from upstream to downstream, the other
one considered different model specifications during different time
periods of the day. Similarly, the authors in \cite{2013_online_arima}
also different configurations of temporal lags in ARIMA model. More
concretely, they firstly applied a hidden Markov model (HMM) model
along with an expectation-maximization (EM) algorithm to evaluate
the traffic state (one of \{Major Accident, Minor Incident, Instability,
Normal Driving\}) in next time slot. After that, the ARIMA models
with different configurations of temporal lags were used to predict
the state of the traffic flow. All these models have improved the
accuracy of the forecasting results compared with ARIMA model. However,
the spatial information was less considered in these models. In this
way, the STARIMA based models \cite{2009_STARIMA,2011_real_starima}
have aroused more and more concern.

\begin{figure}[h]
\begin{centering}
\includegraphics[scale=0.33]{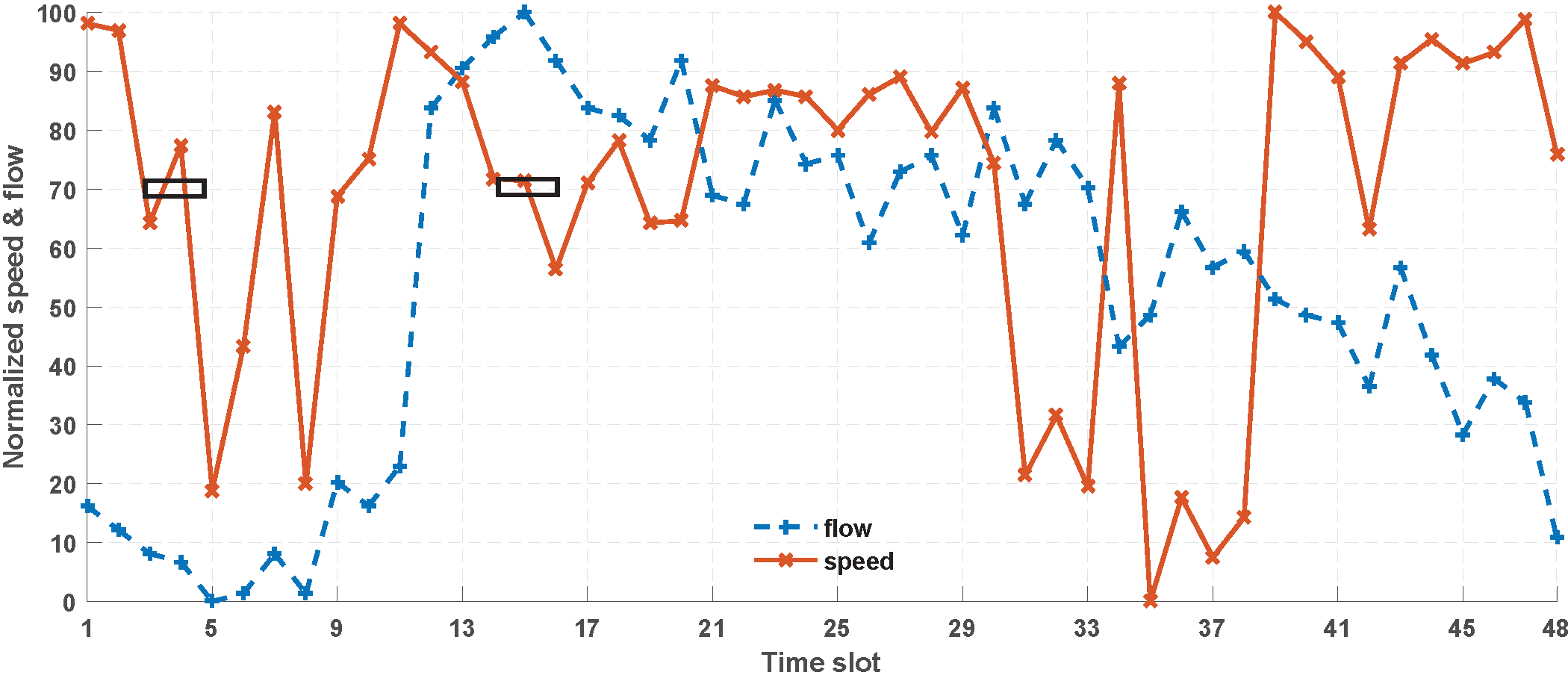}
\par\end{centering}

\caption{The normalized freeway speed and traffic flow data in one day. \label{fig:speed_flow_data}}
\end{figure}

In \cite{2009_STARIMA}, the authors proposed a dynamic STARIMA model
by combining the dynamic turn ratio prediction (DTRP) model and the
STARIMA model. In this paper, a dynamic space weigh matrix is used
to capture different impact of traffic at upstream locations on traffic
at downstream locations. Similarly, the research in \cite{2011_real_starima}
also applied the STARIMA model with the consideration of the dynamic
space weight matrix. Our work distinguish from theirs in that in our
work, the space weight matrices vary with on-peak and off-peak periods
to capture the time-varying correlations of road traffic at different
locations.

From the above related work, we can find that the "``dynamic'' of
a ARIMA or STARIMA model in existing research is often used to indicate
the dynamic of the space weight matrix, the traffic state during different
time periods \cite{2009_graph,2013_road}. However, sometimes it is
not enough to only consider these aspects. For example, Fig.\ref{fig:speed_flow_data}
presents the normalized average speed and normalized flow data collected
at traffic station 6 every 30 minutes in one day. Theoretically, the
weight matrix in time slot 3 (or 4, the left hollow rectangle) and
slot 15 (the right hollow rectangle) should be different since they
are respectively in the off-peak period and peak period. However,
their average speed are the same. This is caused by an inaccurate
evaluation of the time range of peak or off-peak period. Furthermore,
few research considered the relationship between speed and the parameters
(temporal or spatial lag) in STARIMA model. Specially, a great majority
of research use PACF to evaluate temporal lag which easily causes
overfitting problem. Motivated by the above observations, in this
paper we investigate a more efficient method to evaluate these parameters
in STARIMA model with the consideration of spatial information and
the variation of average speed during different time periods.

\section{Data Collection \label{sec:data_method}}

\begin{figure}[tbh]
\begin{centering}
\includegraphics[scale=0.3]{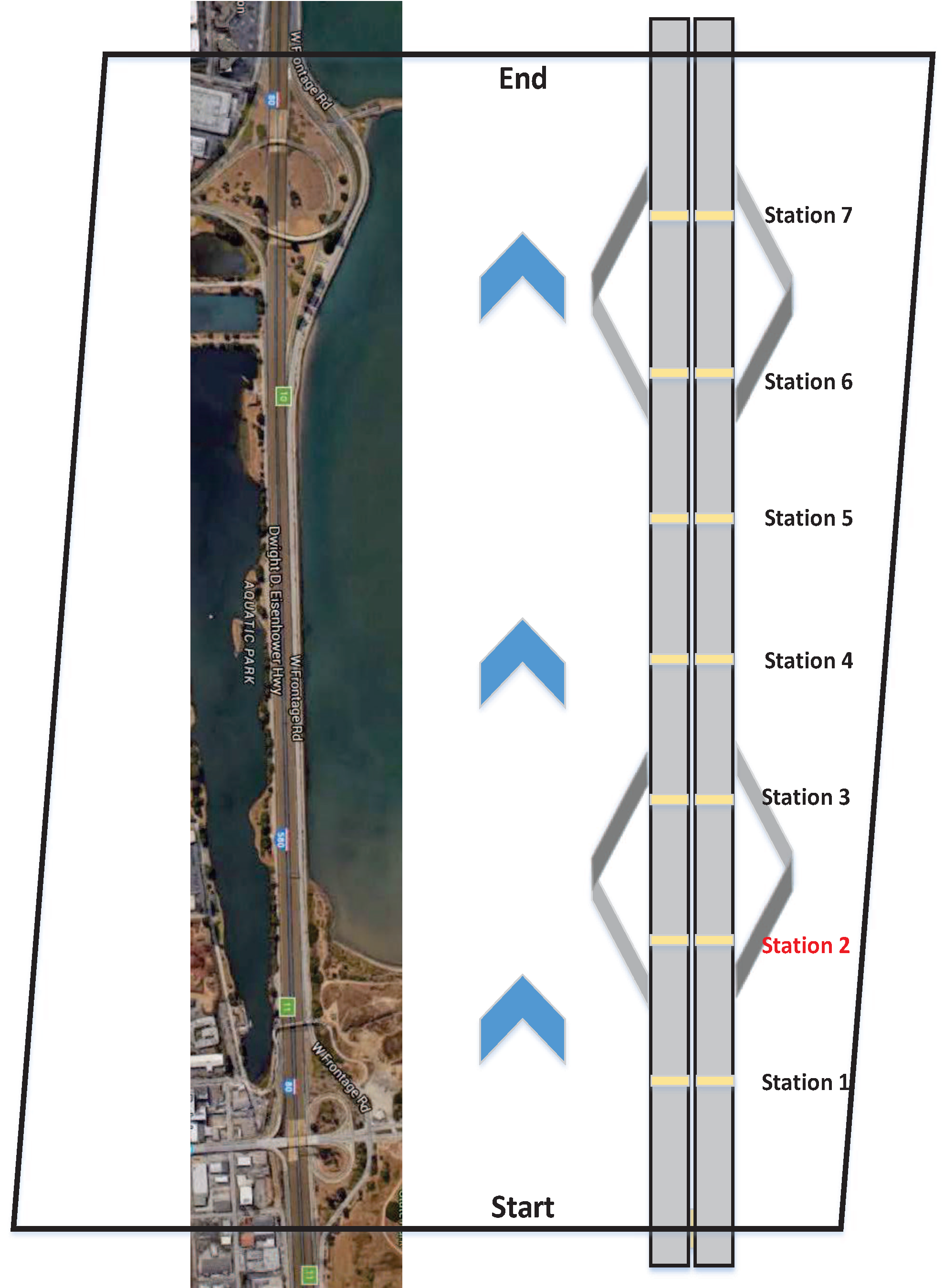}
\par\end{centering}

\caption{The real scenario and topological structure of the I-80 freeway\label{fig:real_scenario_and_topology}}
\end{figure}

\label{sub:data_collection}

In urban environment, the road structure is often complex. Also, the
sensors (such as loop detectors or cameras) \cite{2008_robust,2006_wsn06}
are not deployed at every road. Therefore, it is difficult to obtain
comprehensive data. For simplicity, we only consider highway in this
paper where there are only on-ramp/off-ramps so the traffic condition
is comparatively simpler than that in the urban area. We use data
collected from a segment of Interstate 80 (I-80) freeway located in
Emeryville, California \cite{NGSM}. Available data are collected
every 30 seconds from six traffic stations numbered by 1, 3, 4, 5,
6 and 7 within 10 days. There are two traffic stations for upstream
and downstream traffic respectively. The road topology is shown in
Fig.\ref{fig:real_scenario_and_topology}. Note that there is no data
at station 2 and there are too many interference caused by on or off-ramps
at station 1 and 7. For simplicity, we only use the data collected
at stations 3, 4, 5, and 6. The travel direction is from station 3
to 6.

\section{STARIMA Model with Dynamic Temporal Lag\label{sec:starima_time_lag_variation}}

In this section, we first present a simple way to evaluate the temporal
lag in relation to speed variation. Then we propose a classification
algorithm based on ISODATA by which we can respectively obtain a set
of speed clusters and a set of time period clusters. Finally, we describe
a modified STARIMA model with the varying temporal lag.

\subsection{Temporal Lag with Variation of Speed\label{sub:temporal_lag_variation_speed}}

From Fig.\ref{fig:CCF_3_6_different_time_period}, the temporal lags
with maximal CCF between stations 6 and 3 are different during peak
and off-peak periods. This is attributable to the variation of speed.
Assuming the distance between two detector stations $A$ and $B$
is $L$, and the vehicles keep a stable average speed $\bar{v}$,
then approximately $t=L/\bar{v}$ is needed for vehicles to travel
from $B$ to $A$. In other words, the traffic flow collected at station
$A$ is strongly correlated with that at $B$ $t$ time ago. Thus
the temporal lag with the maximum correlation should be $p=\left[t/\tau\right]$
where $\tau$ is the length of one temporal lag. Note that $L$ depend
on the spatial order $l$ in the STARIMA model and $\bar{v}$ can
often be measured by loop detectors \cite{2000_freeway_speed}. Furthermore,
the advance in telecommunication and electronic technology also brings
a number of new techniques that allows us to estimate the travel time,
e.g. via smartphones. Indeed, the observation discussed in the Introduction
section suggests another novel way to estimate travel time: we can
infer travel time from the correlation of the observed traffic.

In order to validate the above discussion, we further analyze the
results in Fig.\ref{fig:CCF_3_6_different_time_period} by using the
f average speed information at stations 3 and 6, which is collected
in the same day as the traffic flow data used in Fig.\ref{fig:CCF_3_6_different_time_period}.
Specifically, the average speed from station 3 to station 6 between
6:30 am and 8:30 am is 44.45 feet/second. The average speed is 67.05
feet/second between 19 pm and 24 pm. The maximal temporal lag during
these two time periods is respectively 3 and 2 with 30 seconds in
each temporal lag. As $L=v\times p\times\tau$, given $\bar{v}_{1}$and
$\bar{v}_{2}$ during two time periods along with the corresponding
best temporal lag $p_{1}$ and $p_{2}$, we are able to obtain the
following equation according to the theoretical analysis above:

\begin{equation}
v{}_{1}\times p_{1}=v_{2}\times p_{2}\label{eq:speed_times_lag}
\end{equation}
Substituting the data into formulation \eqref{eq:speed_times_lag},
it is easy to find $44.45\times3\approx67.05\times2$. This result
agrees with our theoretical analysis and verifies our speculation
that temporal $p$ is a function of the variation of average speed
$\bar{v}$ in Section \ref{sec:introduction}.

\subsection{The Classification of Speed Data\label{sub:classification_speed}}

An easy way to classify speed data is by dividing into peak time and
off-peak periods. After that, the temporal lag can be calculated using
$p=L/\bar{v}(\pi)$, where $\bar{v}(\pi)$ is the average speed in
time period $\pi,\pi\in$\{peak, Off-peak\}. However an empirical
classification is often prone to error and inaccuracy. Recall the
analysis in Section \ref{sec:related_work}, the evaluation of the
average speed is sensitive to the time range selected for peak or
off-peak period. It is obvious that the speed is not always fast even
during off-peak period from Fig. \ref{fig:speed_flow_data}. Therefore,
in this paper an ISODATA\footnote{More details about the process of ISODATA algorithm are available
in reference \cite{1965_isodata}.} based speed data classification algorithm is developed to deliver
an accurate classification. Using this algorithm, we firstly classify
the speed data collected in each time slot into different clusters.
After that, the time period clusters are confirmed based on the time
slots contained in different speed clusters.

\begin{algorithm}[tbh]
	\begin{algorithmic}[1]
		\State \textbf{Input:} $\boldsymbol{K_{max}}$, $\boldsymbol{n_{min}}$,$\boldsymbol{\sigma_{max}^{2}}$,$\boldsymbol{d_{min}}$,$\boldsymbol{I}$,$\boldsymbol{v}$,$\Delta$
		\State \textbf{Return:} $\boldsymbol{\varGamma}$,$\boldsymbol{\Omega}$
		\State $\boldsymbol{\varGamma}$\textbf{$\leftarrow$ISODATA}($\boldsymbol{K_{max}}$,
		$\boldsymbol{n_{min}}$,$\boldsymbol{\sigma_{max}^{2}}$,$\boldsymbol{d_{min}}$,$\boldsymbol{L}$,$\boldsymbol{v}$)
		\label{algo-isodata}
		\For $\forall\boldsymbol{\varGamma}_{v_{i}}\in\boldsymbol{\varGamma}$
		\label{algo-for-start-1}
		\State $\forall\boldsymbol{\Omega}_{i}={\{T_{i}^{1},T_{i}^{2},...,T_{i}^{K_{i}}\}}$,
		$\emptyset\rightarrow\forall T_{i}^{k}\in\boldsymbol{\Omega}_{i}$
		\State $\forall T_{i}^{1}\leftarrow t_{1},v_{t_{1}}\in\boldsymbol{\varGamma}_{v_{i}}$
		\For $\forall v_{t_{j}}\in\boldsymbol{\varGamma}_{v_{i}}$
		\For $\forall T_{i}^{k}\in\boldsymbol{\Omega}_{i}$
		\If $\exists t\in T_{i}^{k}$ and $t\pm1=t_{j}$
		\State $T_{i}^{k}\leftarrow t_{j}$
		\EndIf
		\EndFor
		\EndFor
		\For $\forall T_{i}^{k}\in\boldsymbol{\Omega}_{i}$
		\State $m=|T_{i}^{k}|$
		\If $m<\Delta$ \label{algo-if-1}
		\For $\forall t_{j}\in T_{i}^{k}$\label{algo-min-start}
		\State $t_{j}\rightarrow min{\{D(t_{j},T_{i}^{\bar{k}})|\forall\bar{k}\in K_{i},\bar{k}\neq k\}}$\label{algo-min-end}
		\EndFor
		\State $\boldsymbol{\Omega}_{i}-{\{T_{i}^{k}\}}$ \label{algo-endif-1}
		\EndIf
		\EndFor
		\State $\boldsymbol{\Omega}\cup\boldsymbol{\Omega}_{i}$ \label{algo-for-end-1}
		\EndFor
	\end{algorithmic}
	\caption{Speed Data Classification\label{alg:Speed-Data-Classification}}
\end{algorithm}

Assuming there is a set of speed data $\mathbf{v}=\{v_{t_{1}},v_{t_{2}},...,v_{t_{n}}\}$
in which $v_{t_{i}}$ is the speed in time slot $t_{i}$. The purpose
here is to confirm a set of speed clusters, denoted as $\mathbf{\varGamma}=\{\varGamma_{v_{1}},\varGamma_{v_{2}},...,\varGamma_{v_{N}}\}$,
where $\forall\varGamma_{v_{i\in N}}\subset\mathbf{v}$ with cluster
center $v_{i}$ and $\forall i,j\in N,\varGamma_{v_{i}}\cap\varGamma_{v_{j}}=\emptyset$.
Based on $\varGamma$, we can obtain another set of time period clusters,
denoted as $\Omega=\{\Omega_{1},\Omega_{2},...,\Omega_{|\varGamma|}\}$,
in which $\forall\Omega_{i}=\{T_{i}^{1},T_{i}^{2},...,T_{i}^{K_{i}}\}$.
In addition, let $T_{i}^{k},k\in K_{i}$ be a set of continuous time
slots, termed as a time range and defined as follows:

\begin{equation}
T_{i}^{k}=\underset{m\geqslant\Delta}{\underbrace{\{t_{j},t_{j}+1,...,t_{j}+m\}}\subset\varGamma_{v_{i}}}
\end{equation}

where $m$ is the number of time slots contained in $T_{i}^{j}$,
$\Delta$ is a threshold defined as the minimal number of time slots
included in a time period. The speed data classification algorithm
is given in Algorithm \ref{alg:Speed-Data-Classification}. In line
\ref{algo-isodata}, the ISODATA algorithm is implemented to get speed
clusters. The time period clusters are obtained from line \ref{algo-for-start-1}
to \ref{algo-for-end-1}. It is worth mentioning that a decision is
made to decide whether $T_{i}^{k}$ belongs to $\Omega_{i}$ by comparing
its capacity and threshold $m$ (from line \ref{algo-if-1} to \ref{algo-for-end-1}).
If $T_{i}^{k}$ does not belong to $\Omega_{i}$, line \ref{algo-min-start}
and \ref{algo-min-end} are implemented to allocate each $t_{j}\in T_{i}^{k}$
to other $T_{i}^{\hat{k}},\hat{k}\neq k$ by the operation $min\{D(t_{j},T_{i}^{\bar{k}})|\forall\bar{k}\in K_{i},\bar{k}\neq k\}$.
$D(t_{j},T_{i}^{\bar{k}})$ is defined as the absolute difference
between speed recorded in time slot $t_{j}$ and the average speed
calculated during time period $T_{i}^{\bar{k}}$, which is presented
in \eqref{eq:absolute_distance}. 

\begin{equation}
D(t_{j},T_{i}^{\bar{k}})=|v_{t_{j}}-\frac{\sum_{t_{\bar{j}}\in T_{i}^{\bar{k}}}v_{t_{\bar{j}}}}{|T_{i}^{\bar{k}}|}|\label{eq:absolute_distance}
\end{equation}

\subsection{$STARIMA(\lambda,p_{\lambda}(v),d,q_{m})$ Model\label{sub:modified_starima_model}}

According to the speed and time period clusters obtained from Section
\ref{sub:classification_speed}, we propose a modified STARIMA model,
denoted as $STARIMA(\lambda,p_{\lambda}(v),d,q_{m})$. The definitions
of parameters $\lambda$, $d$ and $q_{m}$ in this model are the
same as the original STARIMA model, except that the temporal lag $p$
will vary with the spatial order $l$ and the average speed in different
time periods. More precisely, given a time period $T_{i}^{k}\in\Omega_{i}$,
$STARIMA(\lambda,p_{\lambda}(v),d,q_{m})$ is defined as follows:

\begin{equation}
\begin{aligned}(I-\sum_{l=0}^{\lambda}\phi_{l}W_{l}(P_{l}(\bar{v}_{i}^{k})L))(1-L)^{d}Y(t)=\\
(I-\sum_{k=1}^{q}\sum_{l=0}^{m_{k}}\theta_{kl}W_{l}L^{k})\varepsilon_{t}
\end{aligned}
\label{eq:STARIMA_tv_model}
\end{equation}

In \ref{eq:STARIMA_tv_model}, $P_{l}(\bar{v}_{i}^{k})$ is a $N\times N$
vector in which each element $p_{l}^{mn}(\bar{v}_{i}^{k})$ represents
the temporal lag between two station $s_{m}$ and $s_{n}$ with the
spatial order $l$. $p_{l}^{s_{1}s_{2}}(\bar{v}_{i}^{k})_{ij}$ is
calculated by $L(l)/\bar{v}_{i}^{k}$ where $L(l)$ is the distance
between these two stations and $\bar{v}_{i}^{k}$ is the average speed
in time period $T_{i}^{k}$ which is equal to $\frac{\sum_{t_{j}\in T_{i}^{k}}v_{t_{j}}}{|T_{i}^{k}|}$.
Note that when $l=0$, the "``0th order neighbor'' of a station is
itself such that the temporal lag is evaluated with the PACF used
in ARIMA.

\section{Experimental Validation\label{sec:results}}

Based on the data collection introduced in Section \ref{sec:data_method},
we utilize the speed and traffic flow data at stations 3, 4, 5 and
6. At each station, there are 2880 data recorded in one day and the
length of one time slot is 30 seconds. In order to eliminate noise
in the data, we make a "``smooth'' operation by calculating the mean
traffic flow every $x$ data points and regarding it as one data point.
The experimental results are divided into two parts. In the first
part, we provide the speed and time period clusters classified by
our proposed algorithm. For the speed data $\mathbf{v}$, we choose
$x=30$. In the second part, we present the forecast results of traffic
flow in different time periods and stations using $STARIMA(\lambda,p_{\lambda}(v),d,q_{m})$
model. We choose $x=4$ to calculate the mean traffic flow using original
traffic flow data within 2 minutes.

\subsection{The Speed and Time Period Clusters\label{sub:speed_time_cluster}}

Firstly, the configuration of input parameters of algorithm is given
in Table \ref{tab:The-configuration} in which the speed data $\mathbf{v}$
is the results after the smooth operation on the speed data collected
from four stations. With this setting, the smallest length of time
range $T_{i}^{k}\in\Omega_{i}$ is 120 minutes. The speed and time
period clusters classified by Algorithm \ref{algo-isodata} is presented
in table \ref{tab:speed-time-cluster}.

\begin{table}[h]
\caption{The input parameters in Algorithm \ref{algo-isodata}\label{tab:The-configuration}}

\centering{}%
\begin{tabular}{cccc}
\hline 
Parameters & Value & Parameters & Value\tabularnewline
\hline 
$K_{max}$ & 3 & $d_{min}$ & 30\tabularnewline
$n_{min}$ & 5 & $I$ & 10\tabularnewline
$\delta_{max}^{2}$ & 15 & $\Delta$ & 8\tabularnewline
\hline 
\end{tabular}
\end{table}

\begin{table}[h]
\caption{The speed and time period clusters \label{tab:speed-time-cluster}}

\centering{}%
\begin{tabular}{>{\centering}p{1.5cm}>{\centering}p{5.5cm}}
\hline 
Clusters & Values\tabularnewline
\hline 
Speed  & $\varGamma=\{\varGamma_{v_{1}},\varGamma_{v_{2}}|v_{1}=82.15,v_{2}=34.33\}$\tabularnewline
\multirow{3}{1.5cm}{Time period } & $\Omega=\{\Omega_{1},\Omega_{2}|\Omega_{1}=\{T_{1}^{1},T_{1}^{2},T_{1}^{3},T_{1}^{4}\},\Omega_{2}=\{T_{2}^{1},T_{2}^{2},T_{2}^{3}\}\}$\tabularnewline
 & $T_{1}^{1}$={[}0,2am), $T_{1}^{2}$={[}4-6:30am), $T_{1}^{3}$={[}10am-15pm),
$T_{1}^{4}$={[}18:30-24pm{]} \tabularnewline
 & $T_{2}^{1}$={[}2,4am), $T_{2}^{2}${[}6:30-10am), $T_{2}^{3}$=(15-18:30pm{]}\tabularnewline
\hline 
\end{tabular}
\end{table}

\begin{table}[h]
\caption{The temporal lag in $T_{2}^{2}$ and $T_{1}^{4}$ with different spatial
order \label{tab:The-temporal-lag}}

\centering{}%
\begin{tabular}{ccccccccc}
\hline 
\multirow{2}{*}{Day} & \multicolumn{2}{c}{$l$= 3 ($s_{6},s_{3}$)} &  & \multicolumn{2}{c}{$l=2$ ($s_{6},s_{4}$)} &  & \multicolumn{2}{c}{$l=1$ ($s_{6},s_{5}$)}\tabularnewline
\cline{2-3} \cline{5-6} \cline{8-9} 
 & $T_{2}^{2}$ & $T_{1}^{4}$ &  & $T_{2}^{2}$ & $T_{1}^{4}$ &  & $T_{2}^{2}$ & $T_{1}^{4}$\tabularnewline
\hline 
1 & 3/3 & 2/2 &  & 2/2 & 1/1 &  & 1/1 & 1/1\tabularnewline
2 & \textbf{\textcolor{black}{3}}\textbf{\textcolor{red}{/4}} & \textbf{\textcolor{black}{2}}\textbf{\textcolor{red}{/-9}} &  & 2/2 & 1/1 &  & \textbf{\textcolor{black}{1}}\textbf{\textcolor{red}{/-3}} & 1/1\tabularnewline
3 & 3/3 & 2/2 &  & 2/2 & 1/1 &  & 1/1 & 1/1\tabularnewline
4 & 3/3 & 2/2 &  & 2/2 & 1/1 &  & 1/1 & 1/1\tabularnewline
5 & 3/3 & 2/2 &  & 2/2 & 1/1 &  & 1/1 & 1/1\tabularnewline
\hline 
\end{tabular}
\end{table}

There are two speed clusters in $\varGamma=\{\varGamma_{v_{1}},\varGamma_{v_{2}}\}$,
in which $v_{1}$is the cluster center whose value is 82.15 feet/s
and $v_{2}$ is 34.33 feet/s. Based on the speed clusters, one day
is divided into 2 clusters with $\Omega=\{\Omega_{1},\Omega_{2}\}$
in which $\Omega_{1}$includes 4 time ranges and $\Omega_{2}$ includes
3 time ranges. According to such classification, we present the temporal
lag calculated by \eqref{eq:speed_times_lag} (left part of "``/''
in each column) and by the CCF (the right part of "``/'' in each
column) during the time range $T_{2}^{2}$(6:30-10am) and $T_{1}^{4}$(18:30-24pm)
with the spatial lag $l=1,2,3$. The results are shown in Table \ref{tab:The-temporal-lag}.
It reveals an encouraging result that the temporal lags evaluated
by these two methods are the same with the exception of some parts
of the results in day 2. In addition, comparing the temporal lags
evaluated upon different spatial lags, we can find that the temporal
lags during peak and off-peak periods are the same when $l=1$ ($s_{6}$
and $s_{5}$). This is because the temporal lag is less affected by
the speed if two stations are very close.

\subsection{Results of Traffic Flow Prediction}

We choose the traffic flow data in day 3. After the smooth operation,
we predict the traffic flow in one hour (30 time slots), and other
data during each time range $T_{i}^{k}\in\Omega_{i}$ are used for
training $STARIMA(\lambda,p_{\lambda}(v),d,q_{m})$ model (for simplicity,
denoted as $STARIMA(p(v))$). The settings are the same when using
ARIMA and Chaos theory based model (abbreviated as Chaos) \cite{2012_chaos}.
The performance of the forecast is measured by the mean square error
(MSE) and the mean absolute percentage error (MAPE).

\begin{table}[tbh]
\caption{The MAPE/MSE of four stations using \label{tab:mape_mse_starima_time_station}}

\centering{}%
\begin{tabular}{ccccc}
\hline 
$T_{i}^{k}$ & $s_{3}$ & $s_{4}$ & $s_{5}$ & $s_{6}$\tabularnewline
\hline 
$T_{1}^{1}$ & 4.00\%/16.74 & 1.14\%/11.33 & 3.00\%/29.402  & 5.43\%/28.12\tabularnewline
$T_{1}^{2}$ & 11.82\%/46.04 & 6.78\%/63.11 & 1.54\%/11.63 & 6.16\%/39.41\tabularnewline
$T_{1}^{3}$ & 17.85\%/87.90 & 14.34\%/95.31 & 10.99\%/88.97 & 12.62\%/70.69\tabularnewline
$T_{1}^{4}$ & 13.69\%/71.16 & 2.93\%/28.33 & 3.96\%/27.15 & 7.28\%/44.47\tabularnewline
$T_{2}^{1}$ & 12.15\%/ 58.60 & 3.88\%/36.06 & 2.78\%/26.39 & 7.00\%/45.98\tabularnewline
$T_{2}^{2}$ & 14.00\%/75.35 & 4.86\%/59.51 & 2.35\%/29.03  & 8.03\%/51.01\tabularnewline
$T_{2}^{3}$ & 12.20\%/62.79 & 6.61\%/61.82 & 4.24\%/52.79 & 8.68\%/55.56\tabularnewline
\hline 
\end{tabular}
\end{table}

\begin{figure}[tbh]
\begin{centering}
\includegraphics[scale=0.32]{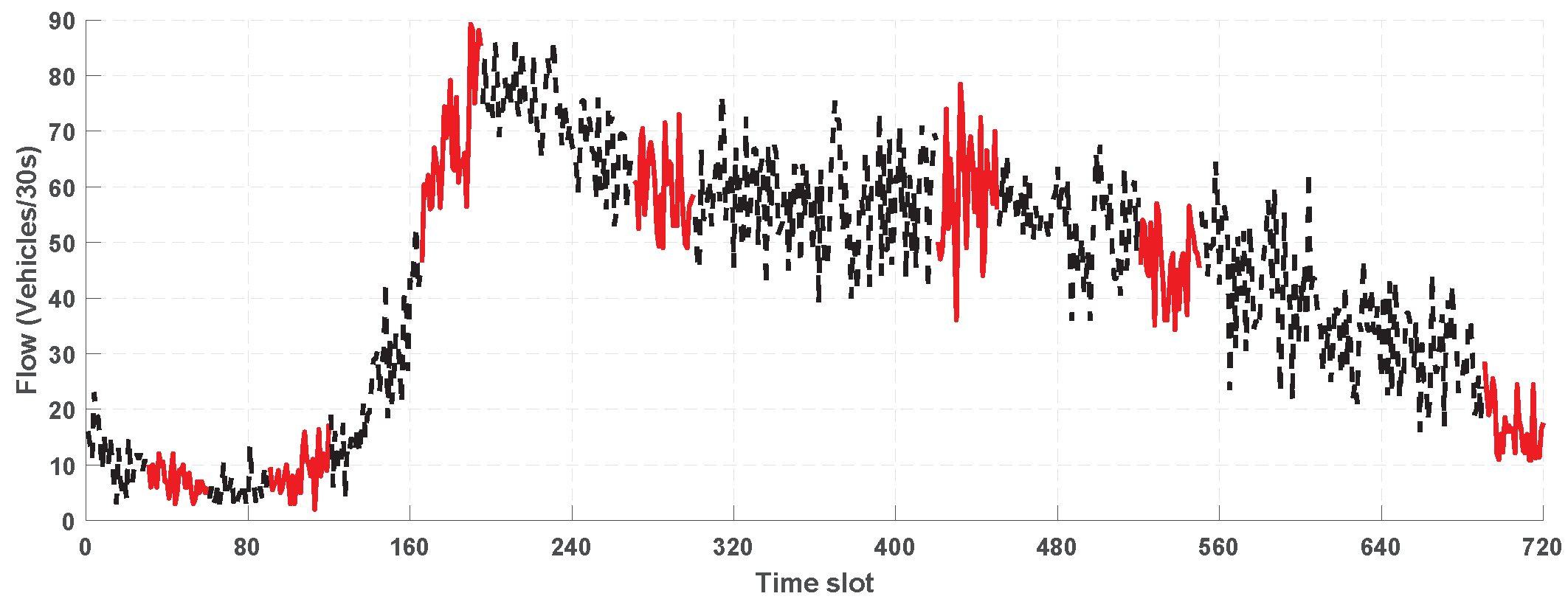}
\par\end{centering}

\caption{The forecasting results for one day at station 6 \label{fig:The-predicted-results-one-day-s6}}
\end{figure}

We provide the MAPE/MSE of four stations using our proposed model
in Table \ref{tab:mape_mse_starima_time_station}. Combining with
Fig. \ref{fig:The-predicted-results-one-day-s6}, we can see that
the forecast results are inspiring. Especially, the MAPE of station
4, 5 and 6 are all below 9\% except $T_{1}^{3}$ (10am-15pm), which
is attributable to the frequent fluctuation of traffic flow during
this time range as shown Fig. \ref{fig:flow_3_6}. As the minimal
time range based on the configuration of $\Delta=8$ is 120 minutes,
it did not capture such frequent fluctuation. In addiction, the MAPE/MSE
of station 3 are higher than other stations because none neighbors
are considered in this paper at this station such that the $\lambda=0$.
In this way, our model is actually similar with a ARIMA model. 

\begin{table}[th]
\caption{The MAPE/MSE of 4 stations using $STARIMA(p(v))$, Chaos, and $ARIMA(2,1,2)$\label{tab:mse_mape_diff_algo_one_day}}

\centering{}%
\begin{tabular}{cccc}
\hline 
St. & $STARIMA(p(v))$ & $Chaos$ & $ARIMA(2,1,2)$\tabularnewline
\hline 
$s_{3}$ & 12.25\%/59.80 & 11.57\%/47.94 & 34.26\%/95.62\tabularnewline
$s_{4}$ & 5.51\%/36.79 & 7.49\%/66.54 & 32.56\%/127.90\tabularnewline
$s_{5}$ & 4.02\%/37.71 & 13.64\%/71.79 & 25.64\%/102.87\tabularnewline
$s_{6}$ & 7.82\%/43.26 & 10.41\%/56.02 & 28.66\%/96.76\tabularnewline
\hline 
\end{tabular}
\end{table}

\begin{table}[th]
\caption{The MAPE/MSE based on forecasting results in 9-10am and 23-24pm at
station 6\label{tab:The-MAPE/MSE-of-two-periods-s6}}

\centering{}%
\begin{tabular}{cccc}
\hline 
St. & $STARIMA(p(v))$ & $Chaos$ & $ARIMA(2,1,2)$\tabularnewline
\hline 
9-10am & 1.28\%/6.52 & 13.01\%/44.70 & 13.11\%/42.74\tabularnewline
23-24pm & 3.38\%/15.55 & 6.12\%/28.55 & 34.82\%/98.11\tabularnewline
\hline 
\end{tabular}
\end{table}

In Table \ref{tab:mse_mape_diff_algo_one_day}, we compare the MAPE/MSE
of one day using $STARIMA(p(v))$, Chaos and $ARIMA(2,1,2)$ at four
stations. Except station 3, all the MAPE/MSE achieved by our model
are better than those of the other two models. Furthermore, in Table
\ref{tab:The-MAPE/MSE-of-two-periods-s6}, we present the MAPE/MSE
of the forecast results of station 6 using these three models, in
which the forecast time ranges are 9-10am and 23-24pm. It can be seen
that the performance of our model is almost coincident with the true
data. And Chaos comes to the second in the prediction during 23-24pm.

\section{Conclusions\label{sec:conclusion}}

Motivated by the observation that the correlation between traffic
at different traffic stations is time-varying and the time lag corresponding
to the maximum correlation approximately equals to the distance between
two traffic stations divided by the speed of vehicles between them,
in this paper, we developed a modified STARIMA model with time-varying
lags for short-term traffic flow prediction. Experimental results
using real traffic data collected on a highway showed that the developed
STARIMA-based model with time-varying lags has superior accuracy compared
with its counterpart developed using the traditional cross-correlation
function and without employing time-varying lags. In an urban environment,
the correlation between traffic tends to be much more intricate. It
is part of our future work plan to develop prediction technique for
urban roads that incorporates the knowledge of the underlying road
topology.


\end{document}